\documentstyle[preprint,aps,prl]{revtex}
\begin{document}
\begin{center}

28 July 1997

\vskip 0.5cm

\Large
{\bf 
Influence of phase-sensitive interaction \\
on the decoherence process\\
in molecular systems}

\vskip 0.5cm

\large

D. Kilin and M. Schreiber

{\sl Institut f\"ur Physik, Technische Universit\"at,
D-09107 Chemnitz, Germany }

\end{center}

\vskip 0.5cm

\noindent
{\bf Abstract}

\noindent
The character of the interaction between an impurity vibrational mode
and a heat bath leads to certain peculiarities 
in the relaxational dynamics of the excited states.
We derive a non-Markovian equation of motion
for the reduced density matrix of this system
which is valid for initial, intermediate and kinetic stages of the relaxation.
The linear phase-sensitive character of the interaction ensures 
the ultrafast disappearance
of the quantum interference of the initially superpositional state
and the effect of classical squeezing of the initially coherent state.
On the other hand, the 
second power
interaction induces a partial conservation of the quantum interference.

\noindent
{\bf Keywords:} decoherence, superposition, 
non-Markovian, vibronic wavepacket
\newline
\noindent
{Corresponding author:}
\begin{minipage}[t]{8.7cm}
Dmitri Kilin \\
Institut f\"ur Physik, 
Technische Universit\"at \\
D-09107 Chemnitz, 
Germany \\
Fax: ++49-371-531-3143 \\
e-mail: d.kilin@physik.tu-chemnitz.de
\end{minipage}

\vskip 3cm

\noindent
{\bf Motivation and model}

\noindent
Time-resolved experimental techniques [1]
sometimes allow to detect quantum superpositional states
in different physical systems [2,3].
Such states provide a basis for the 
possibility of applications like quantum computers [4].
Unfortunately, these interesting states
inevitably  decay due to the coupling with a
heat bath containing many degrees of freedom [5].
Below we consider
how  the character of the coupling with the bath
influences the dynamics of the superpositional
state of the impurity vibrational mode.
Describing this mode as a harmonic oscillator, one writes 
the interaction part of the Hamiltonian as
\begin{equation}
H_I=\hbar \sum_{\xi} K(\omega_\xi) 
\left(
b^+_{\xi} f(a,a^+)+b_\xi f^+(a,a^+)
\right),
\end{equation}
where the function $K(\omega_\xi)$ describes the intensity 
of the interaction between the vibronic mode
and the bath mode operators $b_\xi$.
$f(a,a^+)$ is a function of the vibronic mode operators.
We consider two cases, namely $f(a,a^+)=a+a^+$,
corresponding to the case of the linear phase-sensitive interaction,
and $ f(a,a^+)=a^2$ 
yielding a quadratic
interaction in the rotating wave approximation.
Below we are using ``single''
and ``double'' to indicate the baths 
with these two types of interactions.

\vskip 1cm

\noindent
{\bf Single bath}

The first case describes the processes
when system and bath are exchanging one quantum.
Such a behavior 
is provided by the physical situation
when the majority of the bath modes 
contributing to dephasing and thermalization of the system
have approximately
the same frequency $\omega_\xi$ as the system mode $\omega$.
Applying the formalism of the 
time evolution operator, and restricting to
the second order cumulant expansion [6],
we obtain the non-Markovian master equation [7] 
for the reduced density matrix $\sigma$
of the impurity vibrational mode:
\begin{eqnarray}
\frac{\partial \sigma}{\partial t} 
&=&-i \omega [a^+ a, \sigma] 
\\  &+& ({      \gamma_{n+1}  +\tilde\gamma_n    ^*})[  a   \sigma, a^+ + a] \nonumber
\\  &+& ({       \gamma_{n+1}^* +\tilde\gamma_n })[a^+ + a, \sigma   a^+]   \nonumber
\\  &+& ({\tilde\gamma_{n+1}  +      \gamma_n    ^*})[  a^+ \sigma, a^+ + a] \nonumber
\\  &+& ({ \tilde\gamma_{n+1}^* +      \gamma_n })[a^+ + a, \sigma   a  ] .   \nonumber  
\end{eqnarray}
We obtain four relaxation functions $\gamma$,
which are shown in Fig. 1a.
They originate from the correlations between 
the operators $a(t)$ and $a^+(t)$
of the system mode $\omega$
and the memory kernels of the bath,
like $\left\langle b^+_\xi(0)b_{\xi'} (\tau)\right\rangle$,
which appear in the second order cumulant expansion  
of the evolution operator.
These coefficients are found to be time-dependent.
$\gamma_{n+1}$ and $\gamma_{n}$ describe the situation
when emission-like processes  $\gamma_{n+1}$ prevail over 
the absorption-like processes $\gamma_{n}$,
where $n$ denotes the number of quanta
in the bath modes.
The functions $\tilde \gamma_{n+1}$,
$\tilde \gamma_{n}$ correspond to the reverse situation
and are always small.

The evolution of the different initial states of the system was 
found [6] in analytical form for two cases, namely:
the initial stage of relaxation, when all of the 
relaxation functions
are linear in time,
and for the kinetic stage of the relaxation, when the coefficients
$\tilde \gamma_n$, $\tilde \gamma_{n+1} $ vanish, while
$\gamma_n$, $\gamma_{n+1}$ become constants 
$\Gamma_{1}n_{1}$ and  $\Gamma_{1}(n_{1}+1)$, respectively.
Here $n_{1}=n(\omega)$ indicates 
the number of quanta in the bath mode at the 
system frequency.
The analytical solution is based on the 
generating function formalism [6].
The approaches corresponding to the above stages
are applied to the evolution of the superposition of 
two coherent states $ \left| \alpha \right\rangle $ 
and $e^{i\phi }\left| -\alpha \right\rangle$ 
\begin{equation}
  \left| \alpha ,\phi \right\rangle =N^{-1}\left( \left| \alpha
    \right\rangle +e^{i\phi }\left| -\alpha \right\rangle \right),
  \label{1}
\end{equation}
where $N$ is the normalization constant,
 $ \left| \alpha \right\rangle $
is obtained by displacement
of the vacuum state  $ \left| 0 \right\rangle $
by  $\alpha$ 
as  $\left| \alpha \right\rangle=
\exp(\alpha a^+ -\alpha^* a)\left| 0 \right\rangle $.
The dependence of the probability density $P$
on coordinate $Q$ and time 
is found to consist of classical and interference
parts [7]:
\begin{equation}
P(Q,t)=P_{\rm class}(Q,t)+P_{\rm int}(Q,t).
\end{equation}
The interference part
behaves somewhat different 
under each approach, compare Fig. 1b.
Nevertheless,
energy relaxation of $P_{\rm class}$
and decoherence of $P_{\rm int}$
occur on different time scales,
independent
on the approach applied:
The quantum interference $P_{\rm int}$
disappears faster.
The phase-sensitive character of the
relaxation induces small oscillations
of the broadening of the initially coherent wave packet [7].

\vskip 1cm

\noindent
{\bf Double bath}

When the system interacts with a bath having
the maximum of its mode density 
at twice the oscillator frequency, then
processes occur in which 
the system loses 2 quanta 
and the bath obtains one quantum.
The reverse processes are also allowed. 
To describe such a behavior 
we use $f(a,a^+)=a^2$ in the interaction Hamiltonian ~(1).
The kinetic stage of the evolution of such a system follows
the master equation
\begin{eqnarray}
\frac{\partial \sigma}{\partial t}
&=&-i\omega \left[ a^{+}a,\sigma \right] \label{12a} \\ 
   &+& \Gamma_2 (n_2 + 1)
         \left\{ 
             \left[  
                    a^2
                      \sigma, 
                (a^+)^2
             \right]   
           + \left[ 
                (a^+)^2,
                \sigma
                    a^2 
              \right]
         \right\} \nonumber \\
 &+& \Gamma_2 n_2
         \left\{ 
              \left[ 
                   (a^+)^2 \sigma, a^2 
             \right]   
           + \left[ 
                a^2, \sigma (a^+)^2 
              \right]
         \right\}  
          \nonumber ,
\end{eqnarray}
where $ \Gamma_2 =\pi K^2 g_2 $ is the decay rate of the vibrational
amplitude.
Here, the number of quanta in the bath mode
$n_2=n(2 \omega)$, 
the coupling function
$K=K(2 \omega)$, 
and the density of bath states
$g_{2}=g(2\omega)$ are evaluated at the double frequency 
of the selected oscillator.

Rewritten in the basis of eigenstates  
$\left| n \right\rangle$
of the unperturbed oscillator
this master equation
contains only linear combinations of such terms as
$\sigma_{m,n}=\left\langle m \right|\sigma \left|n\right\rangle$,  
$\sigma_{m+2,n+2}$, and   $\sigma_{m-2,n-2}$.
It ensures, in effect, a possibility to distinguish
even and odd initial states of the system.
The odd excited state $\left| 1 \right\rangle$ 
cannot relax into the ground state $\left| 0 \right\rangle$,
but the even excited state $\left| 2 \right\rangle$ can. 

The evolution of the system 
after preparation under
different initial conditions
was simulated numerically.
The equations of motion of the  density matrix  elements
are integrated using a fourth order 
Runge-Kutta algorithm with stepsize control.
To make the set of differential equations a finite one
we restrict the number 
of levels by $m,n \leq 20$. 

One of the representative examples is the behavior 
of the initially coherent state of the system.
For comparison with the usual behavior we have used
the time dependence of the mean value of the coordinate.
One can see in Fig. 2a that the
relaxation consists of two stages.
The coordinate mean value of 
the usual system decreases with a constant rate
during  both stages.
The same initial value of the system coupled to a 
double bath shows a fast decrement 
in the first stage and almost no decrement
in the second.

The question about the influence of the type 
of the bath on the evolution of the superpositional state
is of special importance.
The simulation was made at a temperature 
of $k_B T = 2 \hbar \omega / {\rm ln}3$,
corresponding to $n(2\omega)=0.5$.
To allow the comparison, we have simulated the evolution
of the same superpositional states twice.
The only difference was the type of the baths, namely
single and double bath.
The coordinate representation 
of the wave packets is presented in Fig. 2b and 2c.
The same value of the relaxation $\Gamma_1=\Gamma_2$ provides, 
however,  different results.
In the system coupled to the single bath 
the quantum interference disappears
already during the first period,
while the  amplitude decreases only slightly.
In the opposite case, the system coupled to
the double bath leaves the interference 
almost unchanged, although
a fast reduction of the amplitude occurs.

Therefore, the second system partially conserves 
the quantum superpositional state.
The experimental investigation of systems displaying
such properties [2] is necessary,
both for extracting typical parameter ranges for theoretical models
 and for practical applications like
quantum computation and quantum cryptography.
\vskip 1cm

\noindent
{\bf Conclusions}

\noindent
The character of the coupling with a heat bath
plays a dominating role in the time evolution of 
different excited 
states of a vibrational mode.
Linear coupling
ensures the ultrafast decay of the 
interference part of the superpositional state.
This result remains true even if different approaches are applied.
The quadratic type of coupling gives 
the same time scales  both for the
amplitude and interference relaxations.
Some initial coherent properties,
like the distinction between even and odd levels,
survive for long time scales.

\vskip 1cm
\noindent
{\bf Acknowledgments}

\noindent
The authors thank DFG for financial support.

\vskip 1cm

\noindent
{\bf References}

\noindent
[1] {\it Femtochemistry: Ultrafast Chemical and Physical Processes 
  in Molecular Systems}, ed. by M. Chergui (World Scientific, Singapore, 1996).

\noindent
[2] J.F. Poyatos,
J.I. Cirac, and P. Zoller,
Phys. Rev. Lett. {\bf 77} (1996) 4728.

\noindent
[3] M. Brune, E. Hagley, J. Dreyer, X. Ma\^itre,
A. Maali, C. Wunderlich, J.M. Raimond,
and S. Haroche,
Phys. Rev. Lett. {\bf 77} (1996) 4887.

\noindent
[4] J.I. Cirac and P. Zoller, 
Phys. Rev. Lett. {\bf 74} (1995) 4091.

\noindent
[5] W.H. Zurek, Phys. Today {\bf 44} (1991) 36.

\noindent
[6] M. Schreiber and D. Kilin, in: {\sl Proc.
2nd Int. Conf. Excitonic Processes in Condensed Matter},
 ed. by M. Schreiber
(Dresden University Press, Dresden 1996) p. 331.

\noindent
[7] D. Kilin and M. Schreiber, Phys. Rev. A (submitted). 

\vskip 2cm

\begin{figure}[p]\centering
  \parbox{7.3cm}
  {\rule{0cm}{0cm}
   \caption[fig2]
  {(a) Time-dependent relaxation coefficients
   $\gamma_{n+1}$ (solid line), 
   $\gamma_n$     (diamonds),
   $\tilde \gamma_{n+1}$ (boxes),
   $\tilde \gamma_n$ (crosses)
   of the master Eq.~(2), 
   calculated for a bath containing
   60 modes in the range $[0,6\omega]$,
   with coupling function
   $K(\omega_\xi) \equiv 1$.
   (b) Time dependence of $P_{\rm int}$
   from Eq.~(4), when the heat bath can be 
   described with only two constants
   $\Gamma_{1}=0.25\omega$,  $n_1=0.4$
   for  $\omega=1$,
   $Q(t=0)=4$, $\phi=0$.
   Crosses: derivation valid for the initial stage of the relaxation,
   solid line: result for the kinetic stage of the relaxation,
   diamonds: corresponding result
   obtained for the linear interaction
   in the rotating wave approximation $f(a,a^+)=a$.
   \label{11}}}
\end{figure}

\begin{figure}[p]\centering
  \parbox{7.3cm}
  {\rule{0cm}{0cm}
\caption[fig2]
{ Difference between double and single baths.
  (a) Evolution of the coherent state:
  Mean value of the coordinate,
  for $Q(t=0)=-2.2$.
  Diamonds: system coupled to
  the single bath, $\Gamma_{1}=0.15\omega$.
  Solid line: system coupled to
  the double bath, $\Gamma_2=0.5\omega$. 
  (b) Time evolution of the superposition of coherent states 
  with initial separation $2Q_0=8$ 
  interacting with single bath (b) for
  $\omega =1$, $\Gamma_1 = 0.005\omega$, 
  $n(\omega)=1.36$.
  (c) Same as (b), but for a double bath with
  $\Gamma_2=0.005\omega$, $n(2\omega)=0.5$.
\label{2}
}}
\end{figure}

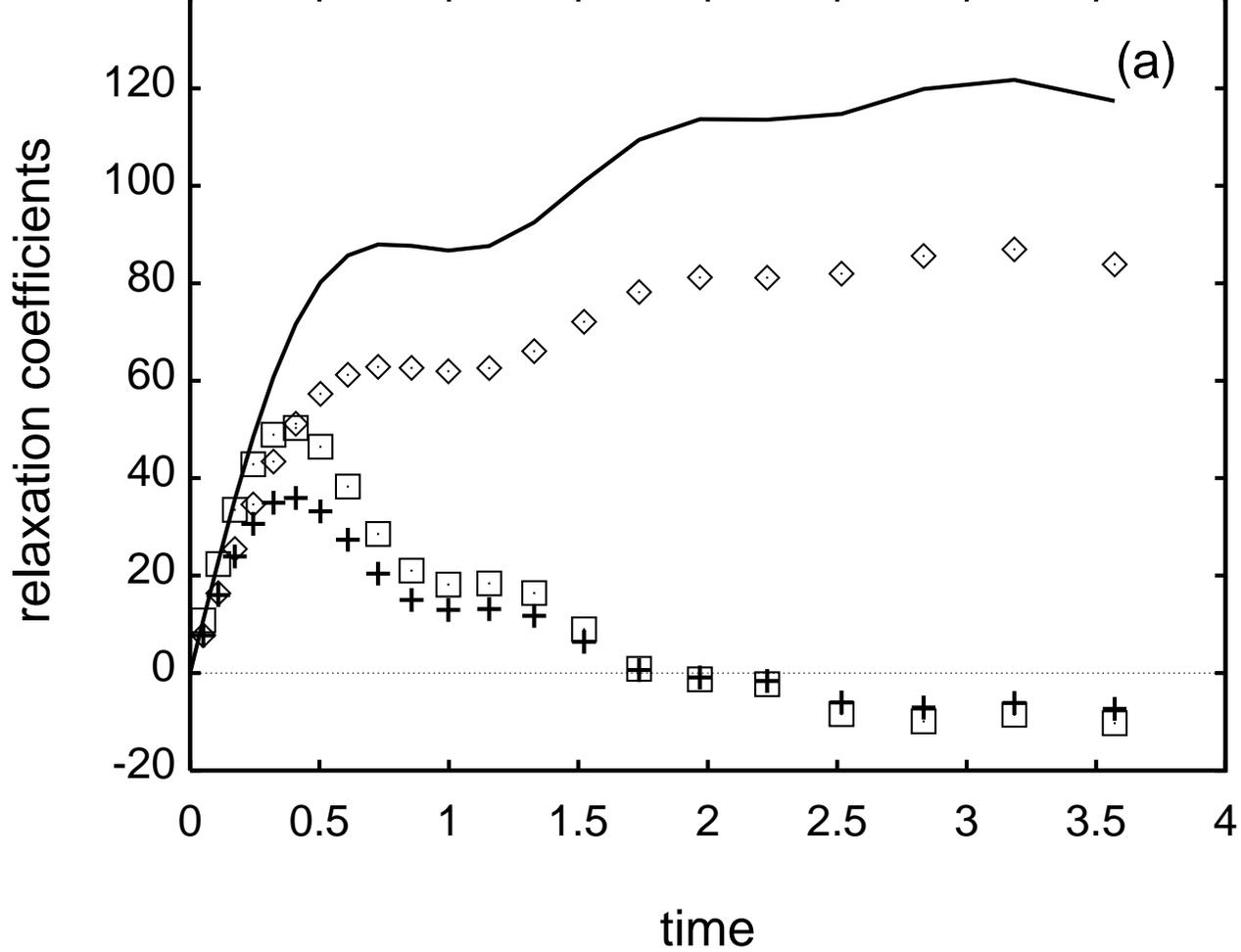
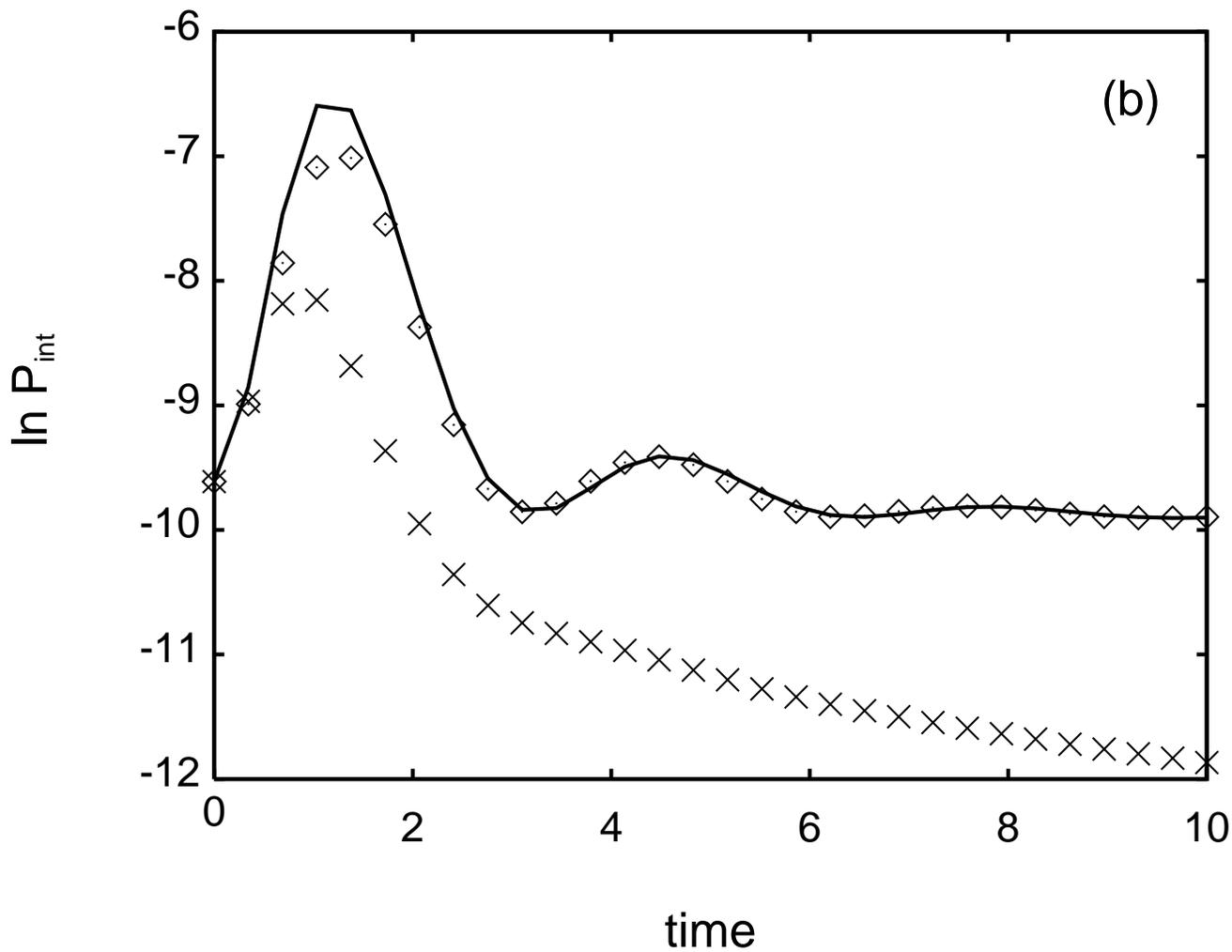
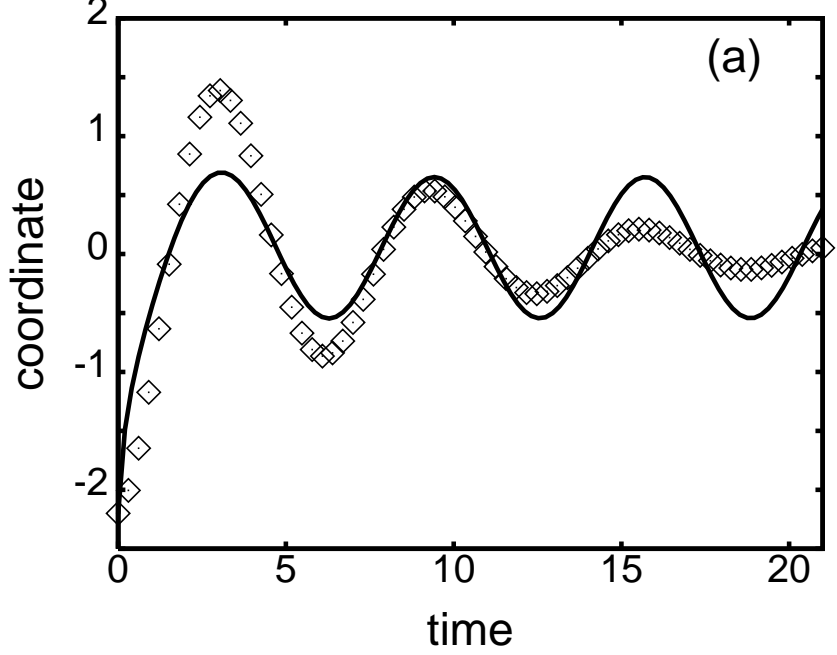
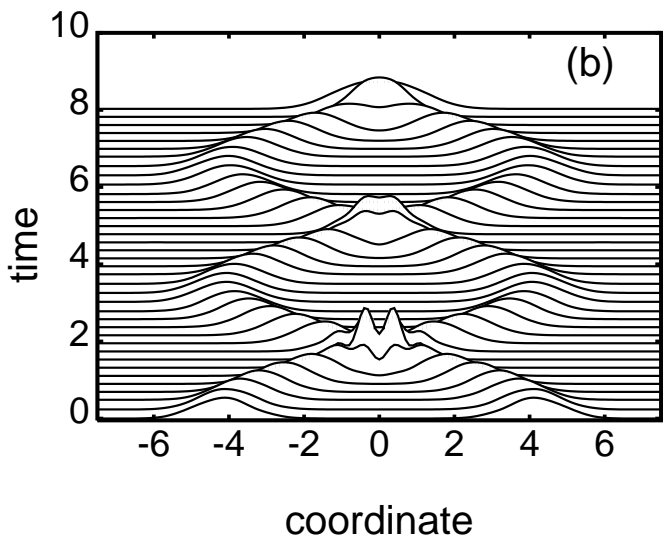
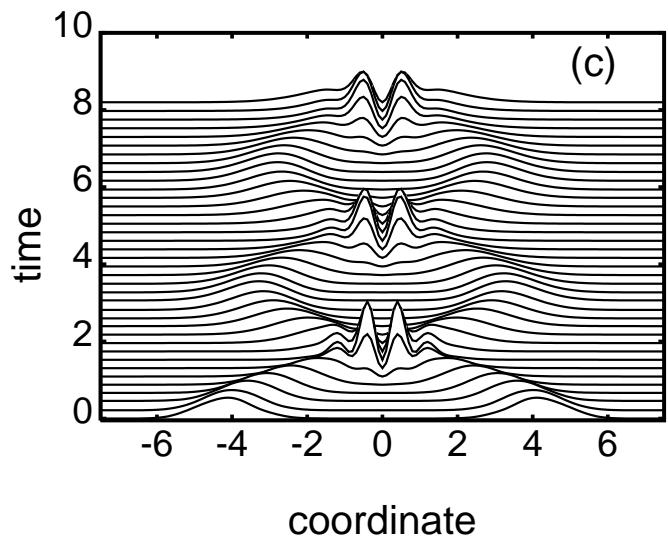
\end{document}